\documentclass{PoS}
\usepackage[english]{babel}
\usepackage{graphicx}
\usepackage{cite}
\usepackage[T1]{fontenc}\usepackage[utf8]{inputenc}
\newcommand{\lyxdot}{.}
\usepackage{epstopdf}
\usepackage{wrapfig}
\usepackage{subcaption}
\usepackage{cleveref}
\newcommand{\ben}{\begin{equation}}
\newcommand{\een}{\end{equation}}

\newcommand{\krsout}[1]{\textcolor{blue}{{}}}
\usepackage{ulem}

\title{Approaching the conformal window: systematic study of the particle spectrum in SU(2) field theory with $N_f = $ 2, 4 and 6.}

\ShortTitle{Approaching the conformal window}
        
     \author{Alessandro Amato,\\ 
     Helsinki Institute of Physics and University of Helsinki\\  E-mail: \email{alessandro.amato@helsinki.fi}\\}
     \author{Teemu Rantalaiho,\\
     Helsinki Institute of Physics and University of Helsinki\\  E-mail: \email{teemu.rantalaiho@helsinki.fi}\\}
     \author{Kari Rummukainen,\\
     Helsinki Institute of Physics and University of Helsinki\\  E-mail: \email{kari.rummukainen@helsinki.fi}\\}
     \author{Kimmo Tuominen,\\
     Helsinki Institute of Physics and University of Helsinki\\  E-mail: \email{kimmo.i.tuominen@helsinki.fi}\\}
     \author{\speaker{Sara Tähtinen}\\
     Helsinki Institute of Physics and University of Helsinki\\  E-mail: \email{sara.tahtinen@helsinki.fi}\\}

\abstract{It is expected that SU(2) gauge theory with $N_f$ fundamental fermions has an infrared fixed point when $N_f$ is between $\sim 6$ and 10. We study the hadron spectrum and scale setting in SU(2) gauge field theory with $N_f=2,4,6$ using hypercubic stout smeared Wilson-clover (HEX) action. The case $N_f=2$ is QCD-like, whereas $N_f=6$ is close to the lower edge of the conformal window. In our study the length scales are determined by using the gradient flow approach.}

\FullConference{The 33rd International Symposium on Lattice Field Theory\\
                 14 -18 July  2015\\
                 Kobe International Conference Center, Kobe, Japan}
\begin{document}

\section{Introduction}

In Technicolor (TC) theories electroweak symmetry is dynamically broken by a chiral condensate of new strong dynamics consisting of SU($N$) gauge field theory with $N_f$ massless fermions. The Higgs sector of the SM then arises as the effective low energy composite description of this new strong dynamics solving the naturalness and fine-tuning problems of the SM. However, the TC paradigm alone does not explain SM fermion masses, which are typically assumed to arise from extended TC (ETC) interactions. Absence of flavour changing neutral current interactions and compatibility with the precision electroweak measurements suggest that the coupling constant should walk, i.e. remain almost constant $g\sim g_\ast$ over a large scale separation. 

%In Technicolor (TC) theories electroweak symmetry is spontaneously broken by a chiral condensate of a new dynamics involving particles that are not present in the Standard Model. TC is SU($N$) gauge field theory with $N_f$ massless fermions, and it solves the fine-tuning problem by substituting SM Higgs scalar with the chiral condensate. However, classical TC scenario does not explain SM fermion masses, but they are regarded in extended TC-theories (ETC). But in ETC, the running of the coupling constant must be slower than in QCD, $i.e.$ it must walk, which means that it remains almost constant $g\sim g_*$ over large scale separation.

At large enough value of $N_f$ the theory looses asymptotic freedom while at low values of $N_f$ the chiral symmetry is broken as in QCD. These are two examples of phases of gauge theories, characterized by the evolution of the gauge coupling, and passage from one another can be studied by  changing the value of $N_f$. Between these two phases, as a function of $N_f$, lies the so called conformal window:  the range of values of $N_f$ for which the $\beta$-function $\beta=\mu\frac{dg}{d\mu}$ of the theory reaches zero at $g^2=g_*^2$. Then the theory has an infrared fixed point (IRFP), leading to scale invariant long distance behaviour. 
%The range of $N_f$ where IRFP exists, is called the conformal window. 
Walking behaviour is expected to occur at the values of $N_f$ direction below the boundary between the conformal window and the phase where the theory breaks chiral symmetry. As the lower boundary of the conformal window lies at strong coupling, lattice simulations are needed to determine the properties of these theories.
%Near the lower edge of the conformal window the walking coupling can be found, whereas below it, chiral symmetry gets broken. IRFP is expected to be found only at strong coupling, where perturbative analysis is not valid and thus lattice simulations are needed.

In this work we study the particle spectrum of SU(2) gauge theory with $N_f=$ 2, 4 and 6 fundamental representation fermions. 
%\krsout{Based on perturbative analysis [any citations?], the theory is expected to be in the conformal window, when $6\leq N_f \leq 10$, but it is not known where the lower edge of the conformal window lies.}
In up to 4-loop perturbation theory, all of these theories have an IRFP \cite{vanRitbergen:1997va}; however, for $N_f=6$ the fixed point appears at very strong coupling.  On the other hand, if one estimates the onset of the chiral symmetry breaking using a ladder approximation, the $N_f=6$ theory is predicted to fall within the conformal window \cite{Appelquist:1996dq,Sannino:2004qp}.
On the lattice, there is clear evidence for chiral symmetry breaking for $N_f \le 4$, and for the existence of an IRFP at $N_f =8$ and $10$ \cite{Karavirta2011,Leino2015}.
However, for $N_f=6$ a clear picture has not yet emerged, with inconclusive and
conflicting results in the literature \cite{Karavirta2011, Bursa2010, Hayakawa2013, Voronov2012, Appelquist2013}.

%\kr{In this project our goal is to make a systematic study of the particle spectrum of the theory when the conformal window is approached and entered, and to resolve whether $N_f=6$ is within the window.  Here we present preliminary results of the particle spectra and scale setting using gradient flow at $N_f=2$, $4$ and $6$.}
%\krsout{for $N_f=$ 2 and 4 there exists clear evidence for chiral symmetry breaking $i.e.$ \cite{Hietanen2014, Karavirta2011}, and $N_f=$ 6 is studied in \cite{Karavirta2011, Bursa2010, Hayakawa2013, Voronov2012, Appelquist2013}, but it is still unclear, whether the theory has IRFP or not. Our goal is to better understand what happens when we are approaching the conformal window, and determine if $N_f=6$ has IRFP.}
%\krsout{The basis of our study is to find how hadron spectrum behaves with small quark masses, and to understand how the scale setting changes as we approach the  conformal window.} We use HEX smeared \cite{Capitani2006} Wilson clover action for fermions, and thin link Wilson plus \krsout{stout} \kr{HEX} link Wilson for gauge fields. We \krsout{have} \kr{use} two lattice sizes: $24^3\times48$, and $32^3\times60$ for small values of $am_Q$. We \krsout{have} \kr{use} 80-200 configurations per parameter set.  \krsout{simulation with lengths 1-2, and the step size is chosen so that the acceptance rate is over 80\%. The scale-setting is studied by using gradient flow approach.}

\section{Mass spectrum}

    \begin{figure}
    \centering
    %\vspace{-1 cm}
    %width=\textwidth,height=5cm,keepaspectratio
    %\includegraphics[scale=0.43]{/Users/sktahtin/plots_lattice/proceedings/Nf2_beta_1\lyxdot 0}
    \includegraphics[scale=0.43]{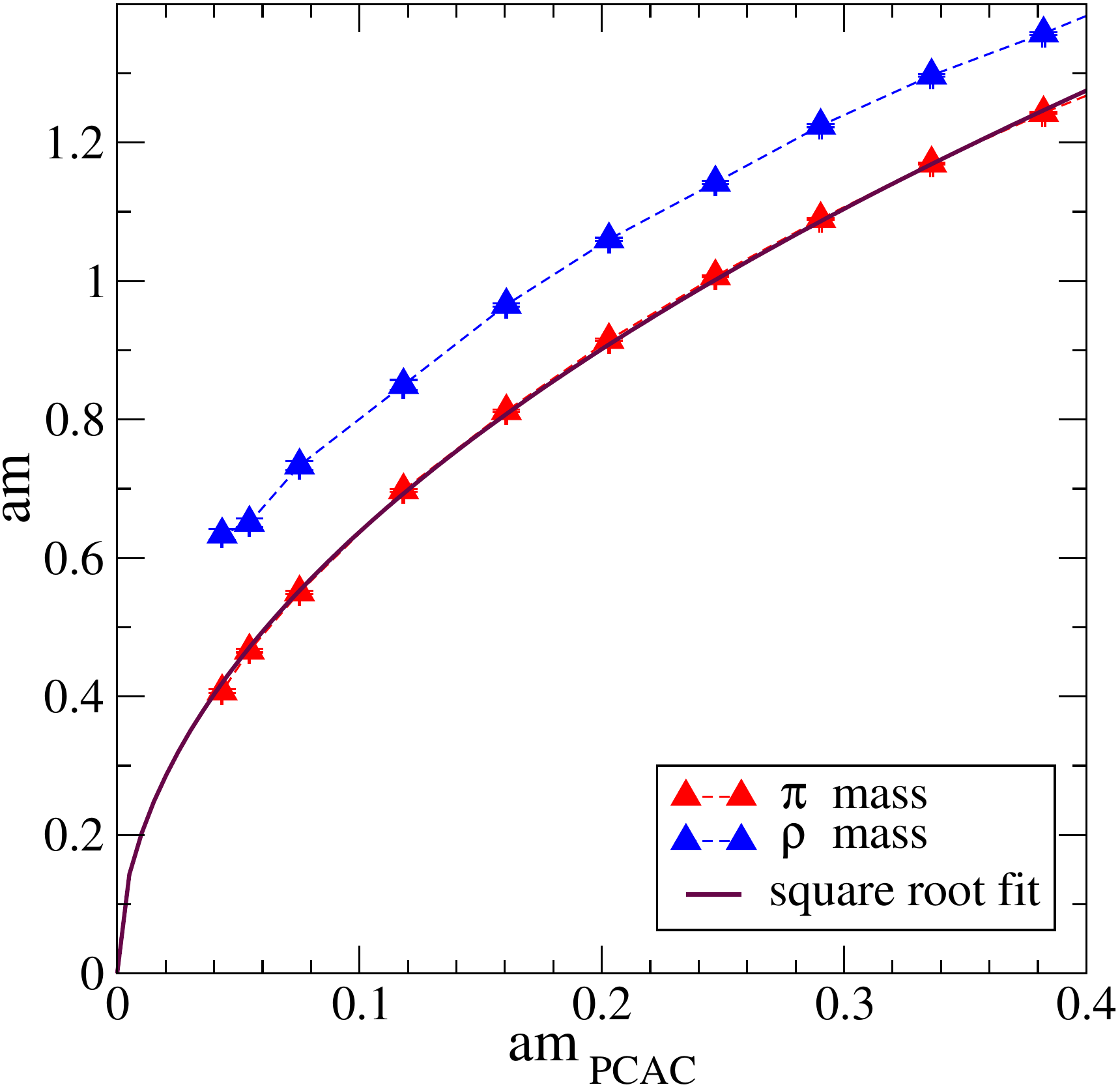}
    \caption{The hadron masses at $N_f=2$ with $\beta=1.0$. 
      The square root behaviour of the pseudoscalar is evident.}
    \label{fig:Nf2}
  \end{figure}

  \begin{figure}
%\centering
  \begin{subfigure}{.5\textwidth}
  \centering
   \includegraphics[scale=0.43]{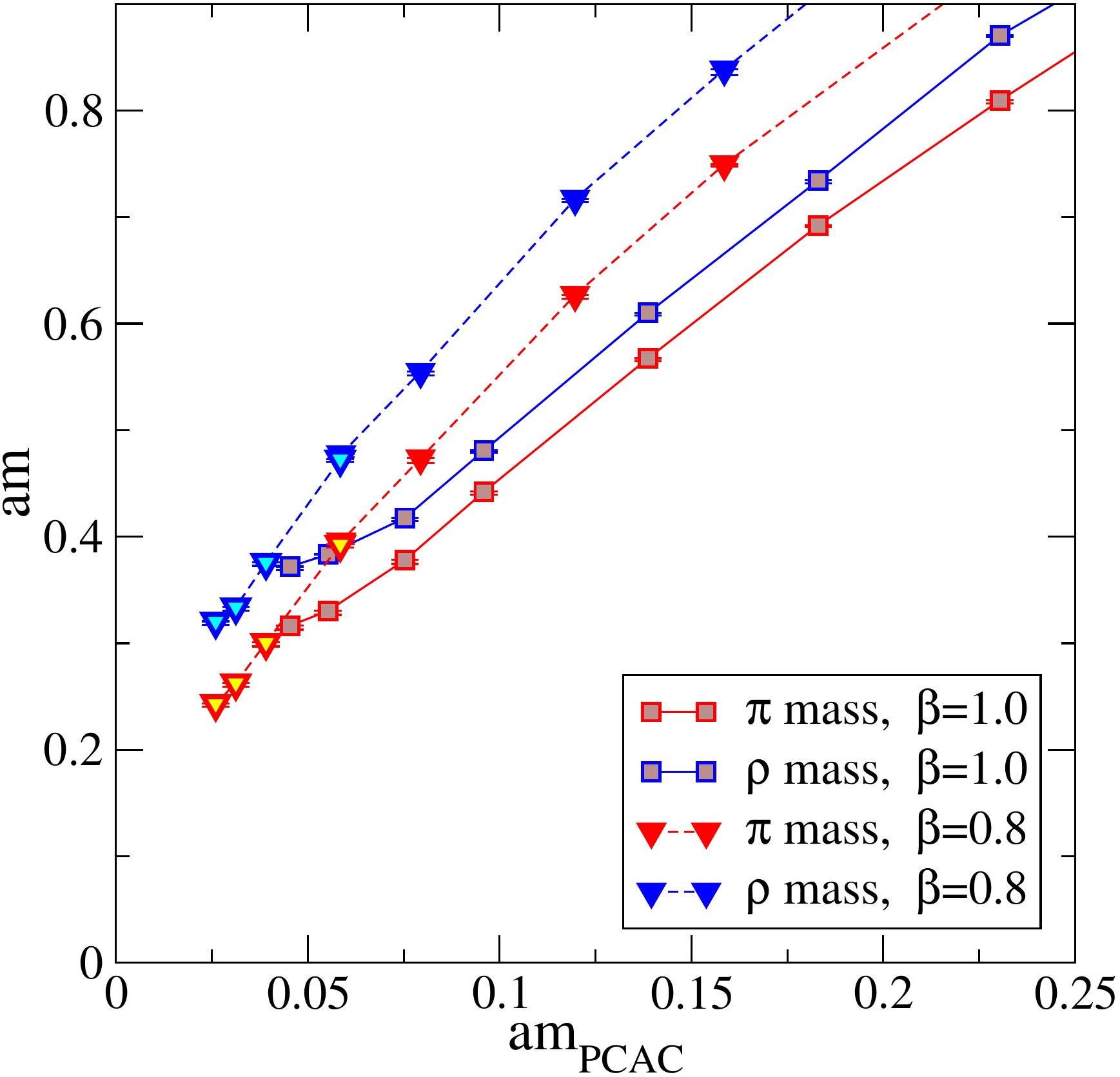}
  %\caption{A subfigure}
  %\label{fig:Nf4_beta1.0}
\end{subfigure}
\begin{subfigure}{.5\textwidth}
  \centering
   \includegraphics[scale=0.43]{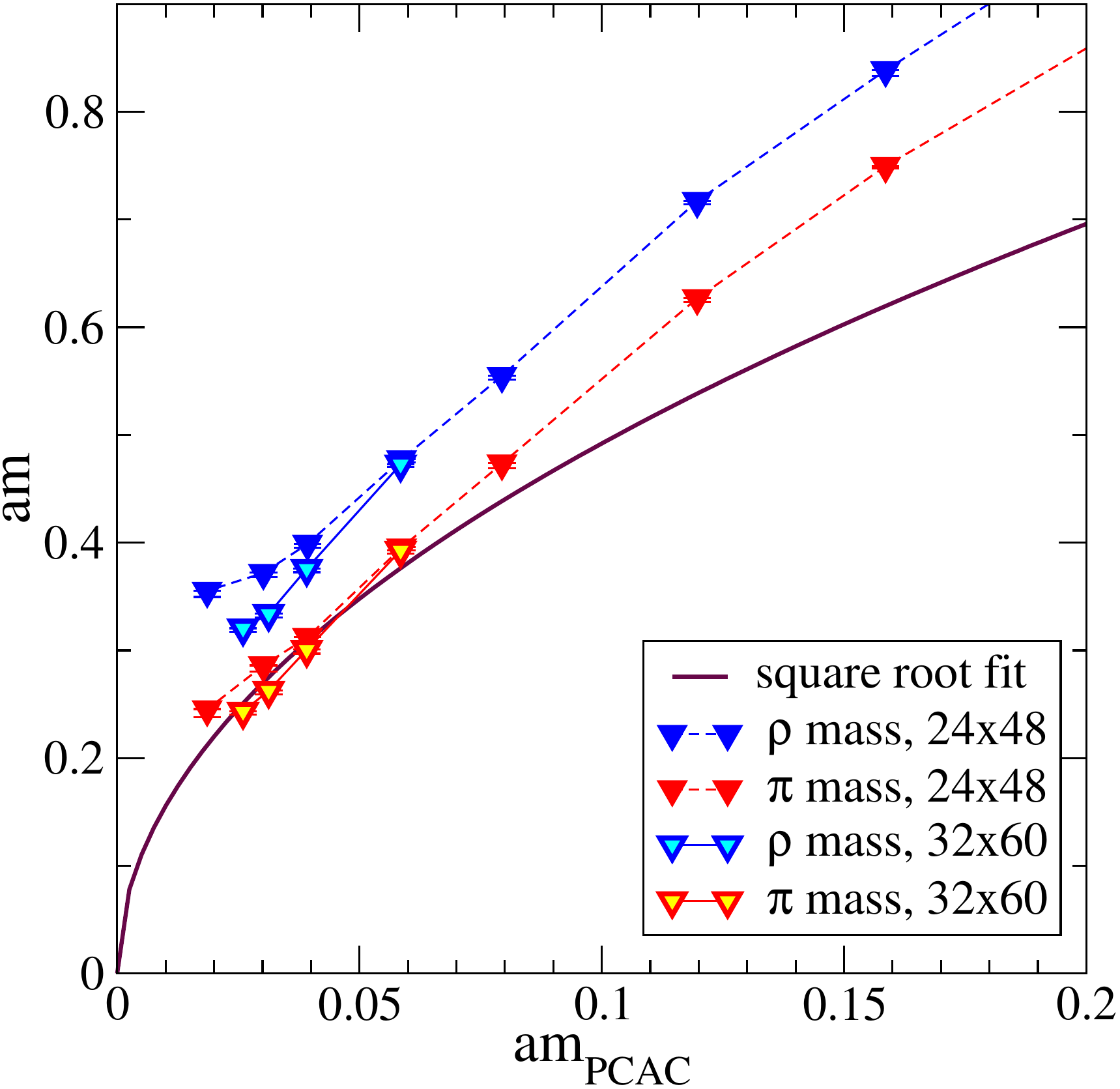}  
  %\caption{A subfigure}
  %\label{fig:Nf4_beta0.8}
\end{subfigure}%
\caption{Left: Pion and rho masses at $N_f=4$ with $\beta= 1$ and $0.8$. The points with yellow center refer to the bigger lattice size ($32^3x60$). Right: the lattice size effect to the hadron masses at $\beta=0.8$.}
\label{fig:Nf4}
\end{figure}
  
  \begin{figure}
%\centering
\begin{subfigure}{.5\textwidth}
  \centering
   \includegraphics[scale=0.43]{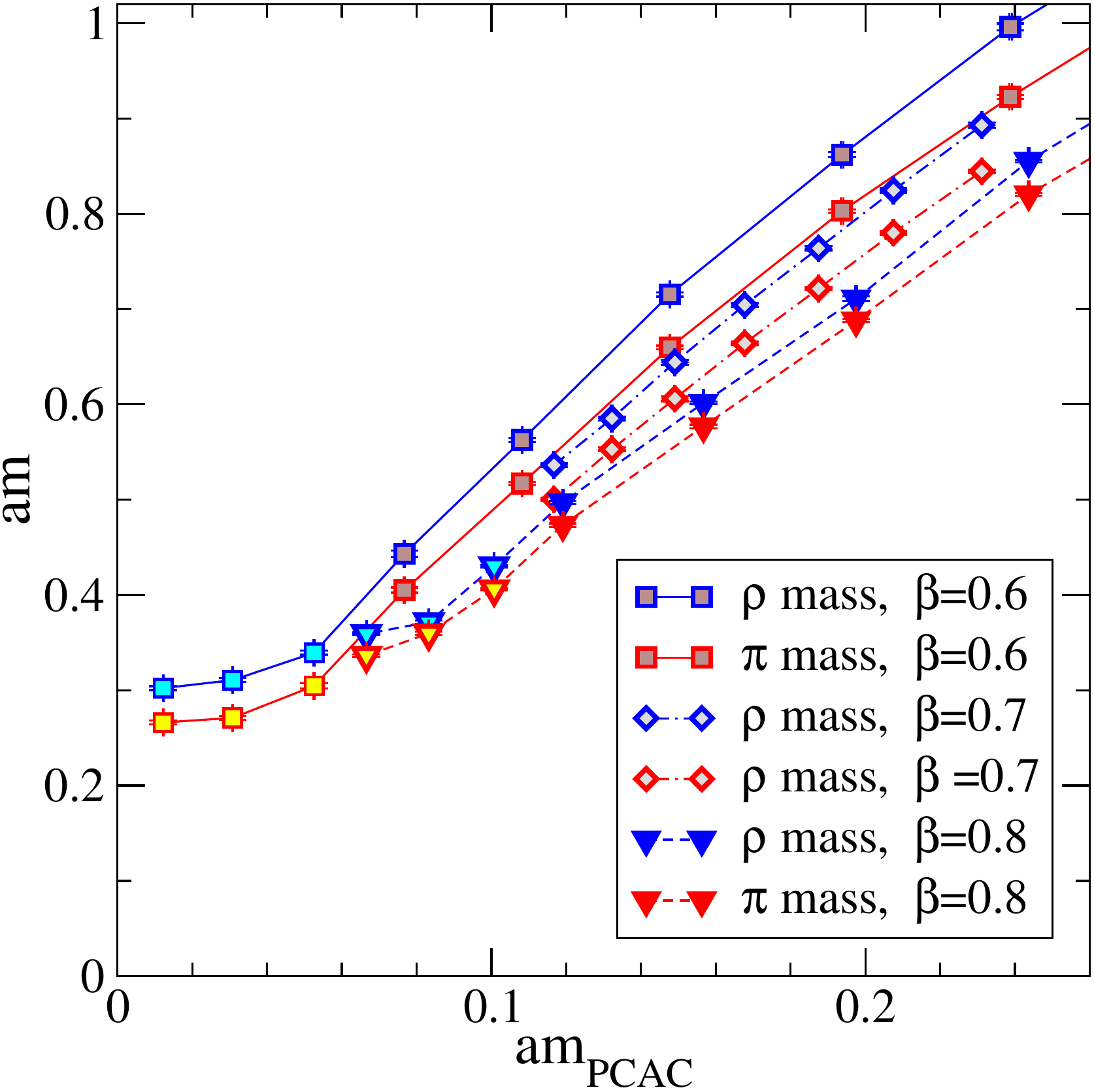}
  %\caption{A subfigure}
  \label{fig:Nf6_beta0.6}
\end{subfigure}
\begin{subfigure}{.5\textwidth}
  \centering
    \includegraphics[scale=0.43]{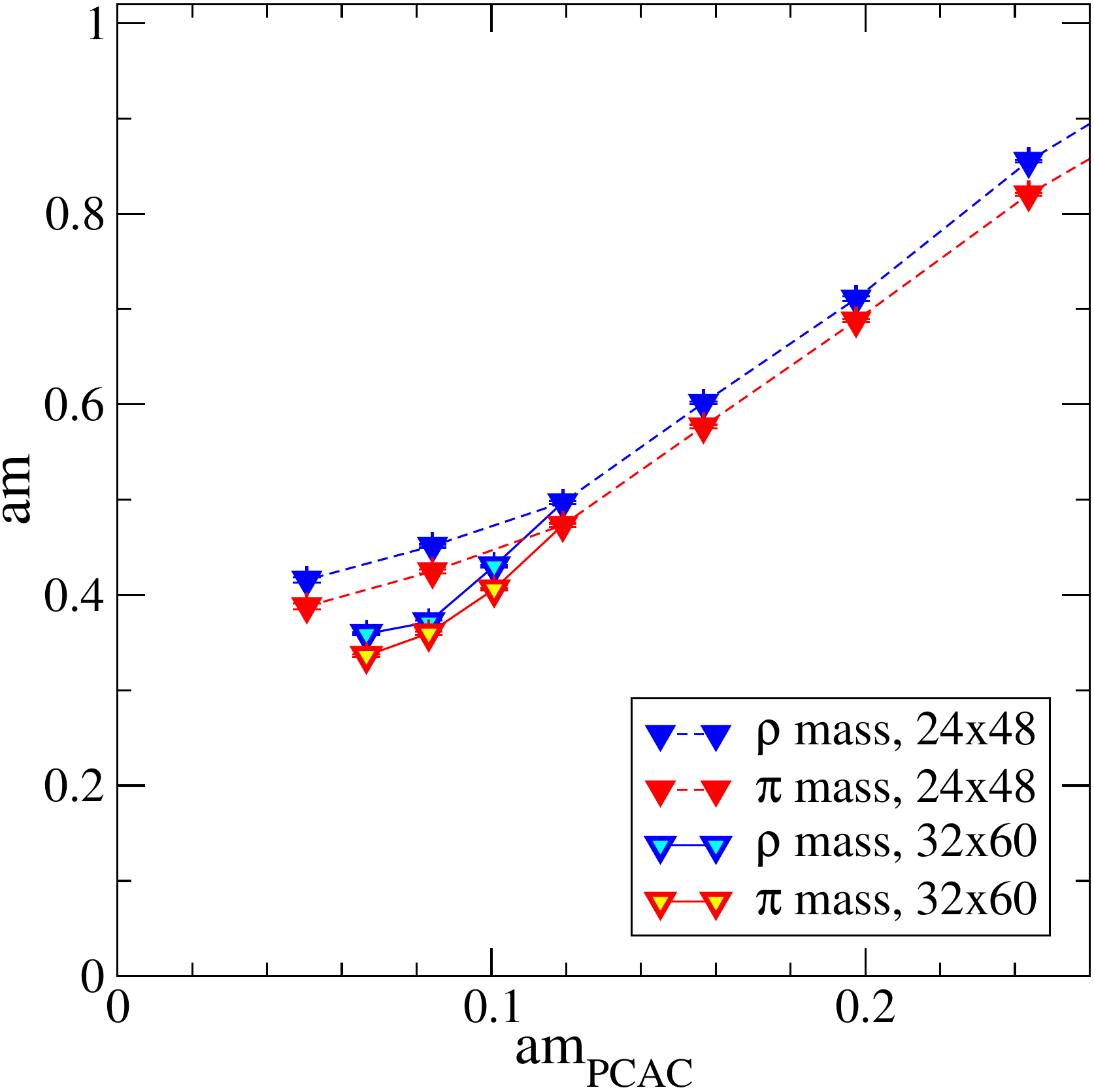}
  %\caption{A subfigure}
  \label{fig:Nf6_beta0.8}
\end{subfigure}%
\caption{Left: Pion and rho masses in $N_f=6$ with $\beta=$ 0.6, 0.7 and 0.8. The points with yellow center refer to the bigger lattice size. Right: The effect of the lattice size on the hadron masses with $\beta=0.8$.}
\label{fig:Nf6}
\end{figure}  
 
%\krsout{In principle, the} The behaviour of the mass spectrum of the theory as the quark mass $m_Q$ becomes small reveals the nature of the theory at long distances. If the theory has QCD-like chiral symmetry breaking, we should observe that the pseudoscalar pions become massless as $m_\pi \propto m_Q^{1/2}$, whereas other bound states remain massive. On the other hand, if the theory has an IRFP, then all states should become massless  $M \propto m_Q^{1/(1+\gamma (g_*^2))}$ as $m_Q \rightarrow 0$. Here $\gamma (g_*^2)$ is the mass anomalous exponent at the IRFP.  In this work, we define the quark masses using the PCAC relation \cite{Luscher1996}.
 
In \cref{fig:Nf2,fig:Nf4,fig:Nf6} the pion and rho masses are presented as functions of the quark mass. At $N_f=2$ the pions behave exactly as expected in QCD-like chiral symmetry breaking scenario, {\em i.e.} pions become massless $\propto \sqrt{m_Q}$.
Because of the familiarity of this scenario we do not attempt to reach very small quark masses.

However, already in the $N_f=4$ case the small quark mass region becomes problematic: the hadron masses in the small $m_{Q}$ region deviate from the expected chiral behaviour by apparently becoming $\sim$ constant and  having a finite intercept as $m_Q \rightarrow 0$. 
This behaviour is due to finite volume effects, as can be observed from the right panel in \cref{fig:Nf4}: the ``levelling off'' happens at heavier quark masses in $24^3\times 48$ lattices than in $32^3\times 60$ ones.   The sensitivity to the finite volume can be explained by the rapid growth in the hadron sizes as $m_Q$ approaches zero, as will be discussed below in connection with the scale setting.  

As expected, the finite volume effects can be avoided to some degree by using
larger volume at fixed lattice spacing (which soon becomes very expensive) or by using coarser lattices.  The lattice spacing dependence is clearly visible on the left panel of \cref{fig:Nf4}, where the levelling off happens at heavier quarks at $\beta=1$ than at $\beta=0.8$.  However, even at the largest volumes and coarsest lattice spacing we are not yet able to verify the expected $m_\pi \propto m_Q^{1/2}$ behaviour.

%\krsout{ This behaviour is also lattice spacing dependent, as even when $32^3\times60$ lattice is used, and the results are not reliable. Thus, if we want to study the region, where the square root behaviour can be found, we should use even bigger lattices, which is, of course, expensive. For $\beta=0.8$ we see hints of a square root behaviour, but with $\beta=1.0$ this ''bending'' behaviour of the hadron masses is even more clear, and happens already with bigger quark masses, so we cannot say whether we see square root behaviour or not. However, from \cref{fig:Nf4} we notice, that pion and rho masses are closer to each, when $\beta$ is larger $i.e.$ physical $m_Q$ are heavier (later explained why). This is expected, since the nature of rho and pion dictates, that $m_\pi$ is about the same as $m_\rho$ in heavy $m_Q$ region.}

The results at $N_f=6$ are still preliminary.
%\krsout{and for $e.g$ $\beta=0.7$ we need more data points.} 
However, we can see that for $N_f=6$ the finite volume effects are even stronger than for $N_f=4$, and it is difficult to reach pion massess which are smaller than $m_\pi \simeq 0.4$ even at coarsest lattices and largest lattice volumes we use.  This indicates that the growth in the size of the hadrons as $m_Q\rightarrow 0 $ is very rapid.

%\krsout{ the "bending" behaviour of the pion and rho masses due to finite volume lattice effects becomes even quicker than in the $N_f=4$ case, and it is worse when the value of $\beta$ is bigger. From this we can deduce that if we want to study what happens in the small PCAC-mass region, we should either use bigger lattice sizes, which is really expensive, or then decrease $\beta$, so use a stronger coupling. As well as in $N_f=4$ case, the pion and rho masses get closer to each other when $\beta$ is increased $i.e.$ when the bare coupling grows.}
  
\section{Scale setting: gradient flow}
  
\begin{figure}
    \centering
    %\vspace{-1 cm}
    %\includegraphics[trim={0cm 0cm 0cm 0cm},scale=1]{/Users/sktahtin/plots_lattice/t0scale-plot-4}
    \includegraphics[trim={0cm 0.5cm 0cm 0cm},scale=1]{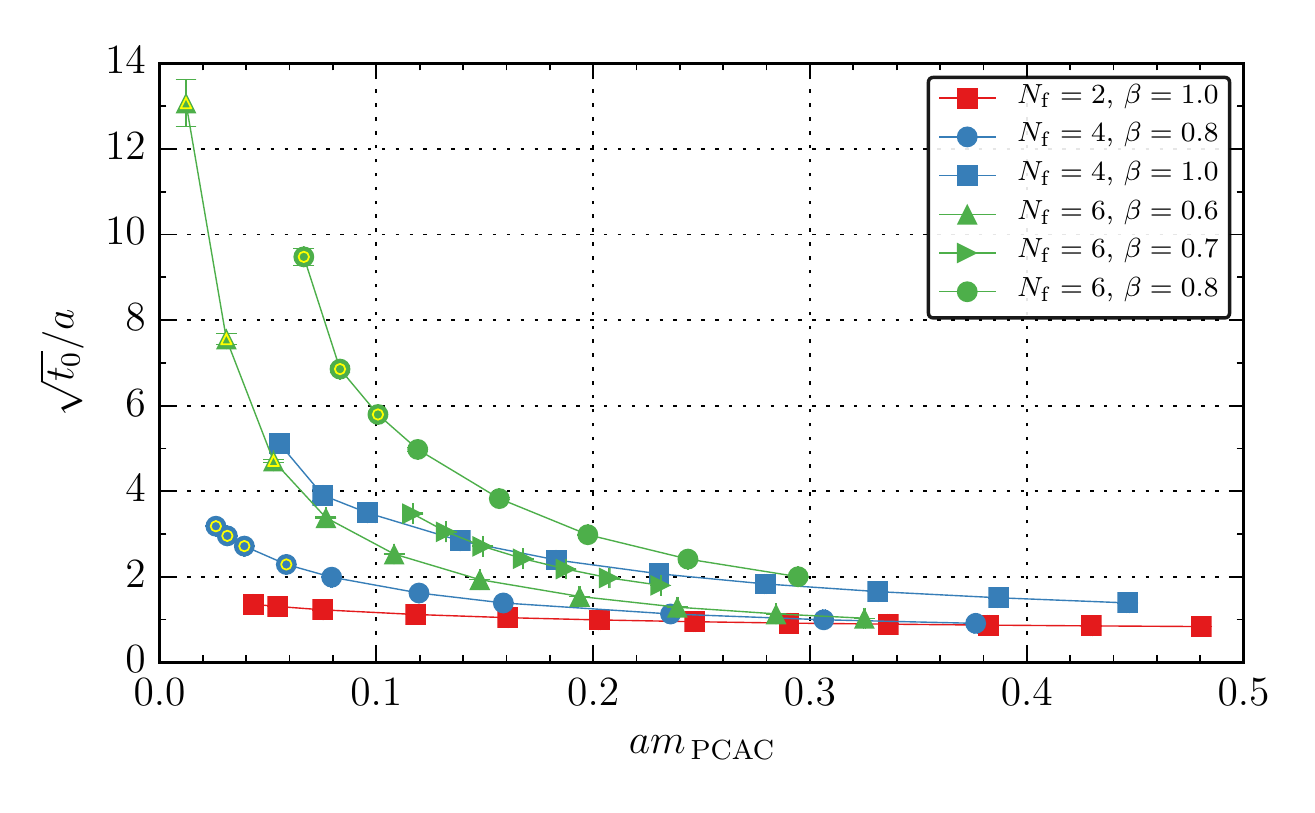}
    \caption{Results of the $\sqrt{t_0}$-scale. Yellow circles and diamonds refer to the results obtained with the bigger lattice.}
        \label{fig:t0}
\end{figure}
  
\begin{figure}
    \centering
    \includegraphics[trim={0cm 0.5cm 0cm 0cm},scale=1]{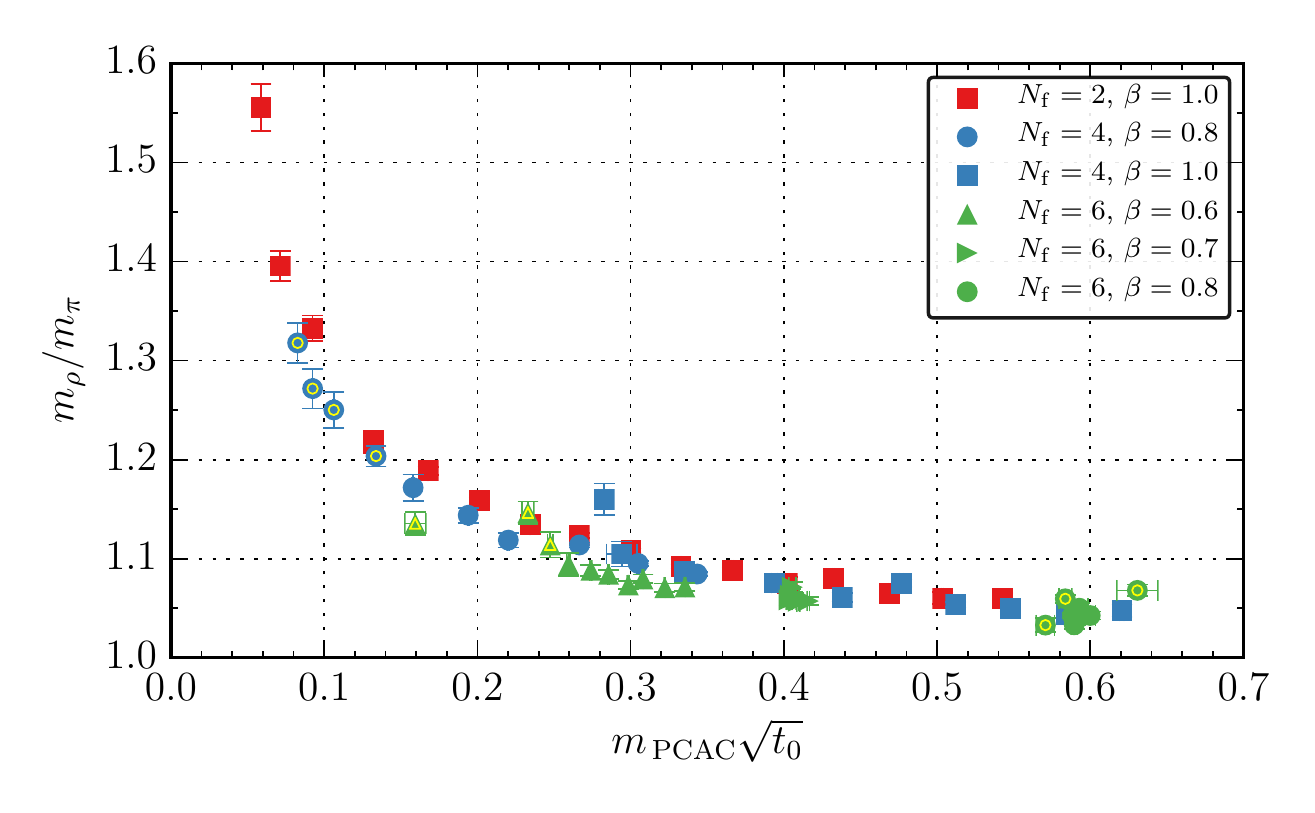}
    \caption{Results in physical units. The results are observed to settle on a universal curve; the underlying reason for this scaling is currently unclear.}
    \label{fig:t0_phys}
\end{figure}

The scale setting procedure in lattice QCD is needed in order to relate the lattice scale to some physically known quantity. 
Popular choices have been rho mass, string tension and the Sommer scale $r_0$ \cite{Sommer1993}, or the Sommer-type scale $r_1$ \cite{Bernard2000}, but presently the gradient flow approach \cite{Luscher2010} has become the preferred one.
Even though it is an artificial way of setting the scale, it is known to be very precise, cheap and straightforward to implement. In order to find the gradient flow length scale $\sqrt{t_0}/a$, one has to solve $t_0$ from
  \ben t_0^2 \left< E(t_0) \right>=0.3, \een
  where $E(t)=\frac14 G^a_{\mu\nu}(t) G^a_{\mu\nu}(t)$ is the %\krsout{continuum-like} 
action density, with $G^a_{\mu\nu}(t)$ the lattice field strength at flow time $t$ \cite{Luscher2010}.  We use here the Symanzik-improved lattice field strength. 
%Sometimes it is also handy to use another, 
A related observable $w_0$ is defined via \cite{Borsanyi2012}
  \ben t\frac{d}{dt}(t^2 \left< E(t) \right>) \vert_{t=w_0^2}=0.3, \een
but in practice these two scales do not differ much. Here we present results using the $t_0$-scale.

In theories in or near the conformal window the scale setting takes another role: if the theory has an IRFP it is scale invariant in the infrared.  Now the scale setting procedure can be used to measure the approach to conformality as the conformal symmetry breaking quark mass $m_Q\rightarrow 0$.  
If the theory has an IRFP we expect that $\sqrt{t_0}/a$ diverges in this limit.
  
In \cref{fig:t0} the scale $\sqrt{t_0}/a$ is presented as a function of $am_Q$. It can be seen that at $N_f=2$ the scale is more or less independent of
$m_Q$, but at $N_f=4$ and $6$ it grows rapidly at small $am_Q$.
While the $N_f=4$ -curves seem to have a finite intecept at $m_q=0$, the
behaviour seen at $N_f=6$ is not obvious.
Because $\sqrt{t_0}/a$ is a measure of the hadron size, strong finite-volume effects at small $m_Q$ can be expected.  
%
%We can also see that we need large enough lattice spacing (small $\beta$), i.e. large enough volume, if we want to reach small $am_Q$.
If we use $\sqrt{t_0}$ to fix the scale, then bigger $\sqrt{t_0}/a$ value  means smaller lattice spacing $a$, and for fixed $a m_{PCAC}$ the quarks are heavier.  In other words, the bigger $N_f$ and the value of $\beta$ are, the harder it is to reach to the small physical quark mass region, because the lattice spacing $a$ gets smaller and smaller. Thus, near the conformal window very strong lattice coupling ($i.e$ small $\beta$) is needed in order to make $m_Q$ small, and to avoid finite volume effects.  
%\krsout{But it is good to keep in mind, that $\beta$ cannot be decreased too much, since then bulk transitions might take place, but we have not seen any hints of that even with $\beta=0.6$.}
  
%\krsout{  It is expected that the physical length scale cannot be set in the small mass region in theories with IRFP, and thus $\sqrt{t_0}/a$ has an infinite value when $am_{PCAC} \rightarrow 0$. From \cref{fig:t0} it seems that for $N_f=$ 2 and 4 $\sqrt{t_0}/a$ has a finite value when $am_{PCAC} \rightarrow 0$, whereas $N_f=6$ grows quicker, so it is  difficult to say if it approaches a finite value or not.}

In \cref{fig:t0_phys} we plot the $m_\rho/m_\pi$ mass ratio against the ``physical'' quark mass $m_Q \sqrt{t_0}$.  Remarkably, all of our measurements  fall on an ``universal'' curve, independent of $N_f$ or the lattice spacing.  For $N_f=2$ the ratio diverges, as expected, as $m_Q\rightarrow 0$.  The $N_f=4$ case follows $N_f=2$, but much more slowly.  Clearly, small $\beta$
(large lattice spacing and large physical volume) is required to reach large $m_\rho/m_\pi$.  
The behaviour of the $N_f=6$ data is interesting: at $\beta=0.8$ and $0.7$ the points remain esentially fixed at $m_Q\sqrt{t_0}=0.6$ and $0.4$, respectively, independent of $m_Q a$, and only at $\beta = 0.6$ the points slowly move towards smaller $m_Q\sqrt{t_0}$ as $m_Q a$ is lowered.  This behaviour may indicate almost conformal behaviour: in a system which becomes conformal as $m_Q\rightarrow 0$ the quark masses are the only physical scale.  Hence $m_Q\sqrt{t_0}$ can be a constant.  However, if the system is not quite
conformal at $m_Q=0$, at light enough quark masses the chiral symmetry breaking behaviour is revealed.  This requires large physical volumes, i.e. small $\beta$.  This may indicate that $N_f=6$ is indeed outside of the conformal
window.  
In the future we will clarify this behaviour further by studying the $N_f=8$ case, which we know has an IRFP  \cite{Ohki2010, Leino2015}.

%\krsout{
%If the theory has chiral symmetry breaking, then $m_{\rho}/m_{\pi}$ must go to infinity as $m_{PCAC} \rightarrow 0$, since $m_{\pi}  \rightarrow 0$.  In \cref{fig:t0_phys} this behaviour can be seen for $N_f=$ 2 and 4, but $N_f=$ 6 is interesting. The $\beta=$ 0.6 grows among the "universal" curve, whereas $\beta=$ 0.7 and 0.8 are stuck at one point. This might mean that there is a fixed point, or it might be that the system has not yet started to grow, since the quark masses are not small enough.
%However, we will next study $N_f=8$, since recent studies suggest that the theory has an IRFP, so it is interesting to know if similar behaviour than in $N_f=6$ is found in the theory with IRFP.}

  \section{Conclusions}
  
  We have initiated a systematic study of the spectrum in SU(2) gauge theory with $N_f=2$, $4$ and $6$ fundamental representation fermions, i.e. when approaching the conformal window.
%with the goal of  Our goal is to see what happens when we are approaching the conformal window. 
In our study, we have focused on the hadron spectrum and the gradient flow scale-setting. 

For the mass spectrum, our central finding is that the finite volume effects are stronger than expected, making it difficult to reach the chiral small quark mass regime even at $N_f=4$.  
In order to reach small quark masses we are forced to use small $\beta$ (strong bare coupling) in order for the physical volume to be large enough.
  
Considering the scale setting, our main observation is that $am_Q$ dependence of the $\sqrt{t_0}/a$ scale becomes stronger as $N_f$ or $\beta$ is increased. 
By plotting the mass ratio $m_\rho/m_\pi$ against $m_Q\sqrt{t_0}$, where
$\sqrt{t_0}$ is the scale obtained from gradient flow, we observed that 
all our measurements fall on an universal curve.  This may indicate that
the chiral symmetry is broken when $N_f\le 6$, i.e. $N_f=6$ is still
outside of the conformal window.

%This means that if the lattice spacing $a$ gets smaller, then $m_Q$ gets bigger, so it is more difficult to get smaller physical masses. Also $N_f=6$ behaves in an interesting way, when when we consider only physical quantities, $i.e.$ plot $m_\pi / m_\rho$ with respect to $m_{PCAC}\sqrt{t_0}$. The datapoints in $\beta=$0.7 and 0.8 are stuck at points $\sim$0.4 and 0.6, whereas $\beta=$0.6 follows the same curve as $N_f=$2 and 4. In order to understand $N_f=6$ better, we will study $N_f=8$, which seems to have IRFP, and perhaps find similarities between $N_f=$ 6 and 8 cases. We also need more datapoints and more values of $\beta$ in $N_f=6$ before we can conclude what is actually happening there.

%LETS NOT REVEAL TOO MUCH HERE  
 %In near future, we will measure the pion and the vector decay constants, and finish up the simulations for $N_f=$ 6 and 8. Later on, we will concentrate on measuring the glueball masses and string tension in these theories, and then we start working on the scalar measurement. 

\section{Acknowledgments}
This work is supported 
by the Academy of Finland grants 267842 and 267286.
S.T. and T.R. acknowledge support from the Magnus Ehrnrooth 
Foundation.  The simulations have been done at the
Finnish IT Center for Science (CSC).

\bibliography{ST_proceedings}

\begin{thebibliography}{10}

\bibitem{vanRitbergen:1997va}
T.~van Ritbergen, J.~A.~M. Vermaseren, and S.~A. Larin.
\newblock {The Four loop beta function in quantum chromodynamics}.
\newblock {\em Phys. Lett.}, B400:379--384, 1997.

\bibitem{Appelquist:1996dq}
Thomas Appelquist, John Terning, and L.~C.~R. Wijewardhana.
\newblock {The Zero temperature chiral phase transition in SU(N) gauge
  theories}.
\newblock {\em Phys. Rev. Lett.}, 77:1214--1217, 1996.

\bibitem{Sannino:2004qp}
Francesco Sannino and Kimmo Tuominen.
\newblock {Orientifold theory dynamics and symmetry breaking}.
\newblock {\em Phys. Rev.}, D71:051901, 2005.

\bibitem{Karavirta2011}
Tuomas Karavirta, Jarno Rantaharju, Kari Rummukainen, and Kimmo Tuominen.
\newblock {Determining the conformal window: SU(2) gauge theory with $N_f =$ 4,
  6 and 10 fermion flavours}.
\newblock {\em JHEP}, 05:003, 2012.

\bibitem{Leino2015}
Viljami Leino, Tuomas Karavirta, Jarno Rantaharju, Teemu Rantalaiho, Kari
  Rummukainen, Joni~M. Suorsa, and Kimmo Tuominen.
\newblock {Gradient flow and IR fixed point in SU(2) with Nf=8 flavors}.
\newblock 2015.

\bibitem{Bursa2010}
Francis Bursa, Luigi Del~Debbio, Liam Keegan, Claudio Pica, and Thomas Pickup.
\newblock {Mass anomalous dimension in SU(2) with six fundamental fermions}.
\newblock {\em Phys. Lett.}, B696:374--379, 2011.

\bibitem{Hayakawa2013}
M.~Hayakawa, K.~I. Ishikawa, S.~Takeda, and N.~Yamada.
\newblock {Running coupling constant and mass anomalous dimension of six-flavor
  SU(2) gauge theory}.
\newblock {\em Phys. Rev.}, D88(9):094504, 2013.

\bibitem{Voronov2012}
Gennady Voronov.
\newblock {Two-Color Schrodinger Functional with Six-Flavors of Stout-Smeared
  Wilson Fermions}.
\newblock {\em PoS}, LATTICE2012:039, 2012.

\bibitem{Appelquist2013}
T.~Appelquist et~al.
\newblock {Two-Color Gauge Theory with Novel Infrared Behavior}.
\newblock {\em Phys. Rev. Lett.}, 112(11):111601, 2014.

\bibitem{Hietanen2014}
Ari Hietanen, Randy Lewis, Claudio Pica, and Francesco Sannino.
\newblock {Fundamental Composite Higgs Dynamics on the Lattice: SU(2) with Two
  Flavors}.
\newblock {\em JHEP}, 07:116, 2014.

\bibitem{Capitani2006}
Stefano Capitani, Stephan Durr, and Christian Hoelbling.
\newblock {Rationale for UV-filtered clover fermions}.
\newblock {\em JHEP}, 11:028, 2006.

\bibitem{Luscher1996}
M.~Luscher and P.~Weisz.
\newblock {O(a) improvement of the axial current in lattice QCD to one loop
  order of perturbation theory}.
\newblock {\em Nucl. Phys.}, B479:429--458, 1996.

\bibitem{Sommer1993}
R.~Sommer.
\newblock {A New way to set the energy scale in lattice gauge theories and its
  applications to the static force and alpha-s in SU(2) Yang-Mills theory}.
\newblock {\em Nucl. Phys.}, B411:839--854, 1994.

\bibitem{Bernard2000}
Claude~W. Bernard, Tom Burch, Kostas Orginos, Doug Toussaint, Thomas~A.
  DeGrand, Carleton~E. DeTar, Steven~A. Gottlieb, Urs~M. Heller, James~E.
  Hetrick, and Bob Sugar.
\newblock {The Static quark potential in three flavor QCD}.
\newblock {\em Phys. Rev.}, D62:034503, 2000.

\bibitem{Luscher2010}
Martin Lüscher.
\newblock {Properties and uses of the Wilson flow in lattice QCD}.
\newblock {\em JHEP}, 08:071, 2010.
\newblock [Erratum: JHEP03,092(2014)].

\bibitem{Borsanyi2012}
Szabolcs Borsanyi et~al.
\newblock {High-precision scale setting in lattice QCD}.
\newblock {\em JHEP}, 09:010, 2012.

\bibitem{Ohki2010}
Hiroshi Ohki, Tatsumi Aoyama, Etsuko Itou, Masafumi Kurachi, C.~J.~David Lin,
  Hideo Matsufuru, Tetsuya Onogi, Eigo Shintani, and Takeshi Yamazaki.
\newblock {Study of the scaling properties in SU(2) gauge theory with eight
  flavors}.
\newblock {\em PoS}, LATTICE2010:066, 2010.

\end{thebibliography}
\bibliographystyle{unsrt}
%\begin{thebibliography}{99}
 % \bibitem{...} ....
%\end{thebibliography}

\end{document}